\documentclass[reprint,amsmath,amssymb,aps,floatfix,]{revtex4-2}
\usepackage{graphicx}
\usepackage{dcolumn}
\usepackage{bm}
\usepackage{xcolor}
\usepackage{physics}
\usepackage{comment}
\usepackage{float}
\begin{document}
\preprint{APS/123-QED}
\title{Time reversal symmetry broken quantum spin hall effect in pseudospin-1 Dirac-Rashba system}
\author{Puspita Parui}
    \email{puspitaparui44@gmail.com}
\author{Bheema Lingam Chittari}%
    \email{bheemalingam@iiserkol.ac.in}
\affiliation{Department of Physical Sciences, Indian Institute of Science Education and Research Kolkata, Mohanpur 741246, West Bengal, India}
\begin{abstract}
The Quantum spin Hall (QSH) phase is conventionally understood to be protected by time-reversal symmetry (TRS). Here, we theoretically investigated the fate of the QSH phase in a pseudospin-1 fermionic $\alpha-\mathcal{T}_3$ system in the presence of a TRS-breaking ferromagnetic exchange field and spin-nonconserving Rashba spin-orbit coupling. Despite broken TRS, the QSH phase survives over a finite parameter regime and is characterised by a non-zero projected spin-Chern number $C_\sigma (\sigma = \uparrow, \downarrow)$, protected by a spin-spectral gap. 
In the absence of Rashba coupling, the QSH phase remains robust up to an $\alpha$-dependent critical exchange field. Rashba SOC qualitatively reshapes the phase diagram by driving transitions into two distinct quantum anomalous Hall (QAH) phases: a $C=2$ phase, irrespective of $\alpha$-values, and a $C=1$ phase for $\alpha \neq 0,1$, which is further identified as a valley-polarized QAH phase arising from a single valley. 
Rotating the magnetization to in-plane gaps out the first-order helical edge states and gives rise to second-order topological insulator (SOTI) phases that host localized corner states in suitable finite geometry. We further identify a topological phase transition between two different SOTI phases, mediated by nanoribbon edge states at an exchange field equal to $\alpha$. 
These results establish spin-resolved topology in a higher pseudospin system as well as the $\alpha-\mathcal{T}_3$ lattice as a versatile platform for engineering and controlling multiple topological phases through magnetic exchange and spin-orbit coupling. 
\end{abstract}
\maketitle
\section{Introduction}\label{sec:intro}
The discovery of topological phases of matter has fundamentally expanded the conventional understanding of condensed-matter systems beyond the framework of symmetry breaking and local order parameters. Among these, the quantum spin Hall effect (QSHE) is one of the most prominent examples, which is a two-dimensional topological insulating phase characterised by a nontrivial $\mathbb{Z}_2$ topological invariant and an odd number of pairs of gapless helical edge states \cite{PhysRevLett.95.146802}. These edge states consist of counter-propagating electrons with opposite spin polarizations and are protected by time-reversal symmetry (TRS), rendering them robust against backscattering from non-magnetic impurities. The QSHE was first theoretically predicted in the Kane-Mele model of graphene, where intrinsic spin-orbit coupling opens a topological bulk gap while preserving TRS \cite{PhysRevLett.95.226801, PhysRevLett.95.146802}. Although the spin-orbit coupling in graphene is too weak for experimental observation, the existence of the QSH phase was subsequently confirmed in HgTe/CdTe quantum wells and later in atomically thin van der Waals materials such as monolayer $\mathrm{WTe}_2$ \cite{bernevig2006quantum,tang2017quantum,shi2019imaging}. Owing to its fundamental significance and potential applications in low-dissipation electronic and spintronic devices, the QSHE continues to serve as a central paradigm in the study of topological quantum matter.\\
A fundamental question concerns the robustness of the QSH phase against perturbations that break TRS. To address this question, various TRS-breaking mechanisms, including exchange fields \cite{PhysRevLett.95.226801, PhysRevLett.95.146802}, magnetic doping \cite{PhysRevLett.101.146802, PhysRevLett.110.266802} and staggered magnetic flux \cite{luo2017time, saha2021eightfold} have been extensively investigated as routes to break TRS in QSH systems. Although TRS is essential for the protection against backscattering of the counter-propagating helical edge states, the QSH phase does not necessarily disappear immediately upon introducing magnetic perturbations \cite{PhysRevLett.107.066602, saha2021eightfold}. In systems with conserved spin, an exchange field may preserve the underlying topological character over a finite parameter range, even though the conventional $\mathbb{Z}_2$ classification ceases to apply \cite{PhysRevLett.107.066602}.
The above picture is further modified when the spin conservation is relaxed. In realistic materials, spin is not always conserved due to Rashba spin-orbit coupling. In the presence of such additional spin-mixing interactions, the competition between magnetic exchange and spin-orbit effects may further drive the system into QAH phases characterised by nonzero Chern numbers and chiral edge states \cite{PhysRevLett.107.066602, PhysRevB.82.161414}. These observations highlight the rich interplay among magnetism, spin-orbit coupling, and topology, motivating the search for new platforms in which such competing effects can be explored.\\
Beyond conventional pseudospin-$1/2$ honeycomb systems, lattices with higher pseudospin degrees of freedom provide an alternative platform for exploring topological quantum phases. In recent years, pseudospin-1 Dirac materials have emerged as an attractive platform for exploring unconventional quantum phenomena, owing to their low-energy excitation described by the Dirac-Weyl Hamiltonian \cite{PhysRevB.34.5208, PhysRevB.96.155301, PhysRevA.83.023609, PhysRevB.81.041410}.
Among them, the $\alpha-\mathcal{T}_3$ lattice has attracted considerable attention owing to its unique electronic structure and tunable geometric properties. The lattice interpolates continuously between graphene ($\alpha=0$) and the dice lattice ($\alpha=1$) through the parameter $\alpha$, while supporting an additional dispersionless flat band intersecting the two linearly dispersing Dirac bands \cite{PhysRevLett.112.026402, PhysRevB.92.245410}.
The Berry phase in the $\alpha-\mathcal{T}_3$ lattice varies smoothly from $\pi$ $(\alpha=0)$ to 0 $(\alpha=1)$ with $\alpha$, leading to unusual transport and topological responses \cite{PhysRevB.92.245410, li2022novel, PhysRevB.107.085408, PhysRevB.111.024416}. 
The interplay between the flat band, Dirac-like dispersive bands, and the tunable Berry phase gives rise to a variety of unconventional phenomena, including $\alpha$-dependent quantum Hall \cite{li2022novel}, quantum spin Hall \cite{PhysRevB.103.075419}, and quantum anomalous Hall phases \cite{PhysRevB.101.235406, PhysRevB.99.205429}. 
The $\alpha-\mathcal{T}_3$ lattice is long been proposed to be realized experimentally, using three pairs of counter-propagating laser beams in a cold atom setup \cite{PhysRevA.80.063603} as well as in transition metal oxide heterostructure such as $\mathrm{SrTiO}_3/\mathrm{SrIrO}_3/\mathrm{SrTiO}_3$ grown along $(111)$-direction \cite{PhysRevB.84.241103}. Furthermore, it has been shown that critically doped $\mathrm{Hg}_{1-x}\mathrm{Cd}_x\mathrm{Te}$ can be effectively mapped onto an $\alpha-\mathcal{T}_3$ model with $\alpha=1/\sqrt{3}$ \cite{PhysRevB.92.035118}. More recently, direct experimental evidence of the Dice lattice flat band has been reported in the 2D van der Waals electride YCl, where the lattice geometry is formed by interstitial anionic electrons (IAEs) \cite{4sq5-x2jy}. 
Beyond its topological properties, the $\alpha-\mathcal{T}_3$ lattice has also been shown to exhibit a wide range of unconventional physical phenomena, including enhanced thermoelectric properties \cite{alam2019enhancement, PhysRevB.108.115141, tamang2026spin}, optical responses \cite{PhysRevB.92.245410, PhysRevB.94.125435, PhysRevB.95.035414, PhysRevB.99.045420, PhysRevB.100.035440, PhysRevB.105.155405, PhysRevB.106.115143}, valley polarized transport \cite{PhysRevB.96.045418,li2021valley}, anomalous magneto transport \cite{PhysRevB.107.245150, PhysRevB.102.235414, biswas2016magnetotransport}, higher Chern insulating phases \cite{PhysRevB.107.035421, PhysRevB.101.235406, PhysRevB.109.165118, PhysRevB.104.235115}, and Floquet dynamics \cite{PhysRevB.99.205429, PhysRevResearch.4.033194, PhysRevB.104.174308, benhaida2025topological, PhysRevB.111.045406}.
Although the topological properties of the $\alpha-\mathcal{T}_3$ lattice in the presence of intrinsic spin-orbit coupling have been extensively investigated, revealing $\alpha$-dependent QSH phases \cite{PhysRevB.103.075419, PhysRevB.111.045406}, and the effect of Rashba SOC and out-of-plane ferromagnetic exchange field has also been explored in the context of magnetic and non-magnetic staggered potential effects \cite{4sq5-x2jy}, a systematic understanding of the interplay among intrinsic spin-orbit coupling, exchange coupling, and Rashba spin-orbit coupling is still lacking.
In particular, it remains an open question how the $\alpha$-dependent QSH phase evolves when time-reversal symmetry is broken by an exchange field and spin conservation is relaxed by Rashba SOC. Resolving this issue is essential for understanding the stability of the QSH phase, the emergence of new topological phases, and the role of tunable lattice parameter $\alpha$ in governing the resulting topological phase diagram. \\
In this work, we investigate the fate of the $\alpha$-dependent quantum spin Hall phase in the Kane-Mele $\alpha-\mathcal{T}_3$ lattice under the simultaneous breaking of TRS and spin conservation. Starting from a general exchange field, we systematically investigate the effects of both out-of-plane and in-plane magnetization on the topological properties. 
By analysing the bulk band structure, topological invariants, and edge-state spectra, we demonstrate how time-reversal symmetry breaking drives transitions between distinct topological phases. Furthermore, we show that the resulting phase diagram is strongly controlled by the lattice parameter $\alpha$, revealing a nontrivial interplay among pseudospin structure, magnetization, and spin-orbit coupling. Our results highlight the $\alpha-\mathcal{T}_3$ lattice as a versatile platform for engineering and controlling topological phases beyond those accessible in conventional honeycomb systems.\\
The remainder of the paper is organised as follows. In Sec.~\ref{sec:model_hamiltonian}, we present the model Hamiltonian describing the system under consideration. The numerical results are discussed in Sec.~\ref{sec:results}, where Sec.~\ref{sec:section_A} presents the bulk band-gap phase diagrams with their topological characterisations and Sec.~\ref{sec:section_B} examines the edge-state spectra for the relevant topological phases in the presence of out-of-plane magnetization. The effects of in-plane magnetization on the QSH phase are discussed in Sec.~\ref{sec:section_C} in the context of second-order topological insulator (SOTI) modes and their real-space spectra. Finally, we summarise our findings in Sec.\ref{sec:conclusion}.

\section{Model Hamiltonian}\label{sec:model_hamiltonian}
The tight-binding model for the QSH $\alpha-\mathcal{T}_3$ Kane-Mele-Rashba model with Rashba SOC and ferromagnetic exchange coupling with arbitrary magnetization orientation can be written as,

\begin{equation}
    H = H_\text{KM} + H_\text{R} + H_\text{M}
\label{eq:full_hamiltonian}
\end{equation}
Where,
\begin{equation}
\begin{aligned}
    H_\text{KM} &= -t \sum_{\langle i,j \rangle, s} c_{i,s}^\dagger c_{j,s} - \alpha t \sum_{\langle j,k \rangle, s} c_{j,s}^\dagger c_{k,s}\\
    &\quad + \frac{i}{3\sqrt{3}}\sum_{\langle\langle i,j \rangle\rangle} \lambda_I^i \nu_{ij} c_{i,s}^\dagger [{\bm{\sigma}}_z]_{s,s^\prime} c_{j,s^\prime} \\
    &\quad + \frac{i\alpha}{3\sqrt{3}}\sum_{\langle\langle j,k \rangle\rangle} \lambda_I^i \nu_{jk} c_{j,s}^\dagger [{\bm{\sigma}}_z]_{s,s^\prime} c_{k,s^\prime}
\end{aligned}
\label{eq:km_hamiltonian}
\end{equation}

\begin{equation}
\begin{aligned}
    H_\text{R} &= \frac{2i\lambda_R}{3} \sum_{\langle i,j\rangle,s,s^\prime} \hat{\bm{e}}_z \cdot (\bm{\sigma}_{s,s^\prime} \times \bm{d}_{i,j}) c_{i,s}^\dagger c_{j,s^\prime} \\
    & \quad + \frac{2i\alpha\lambda_R}{3} \sum_{\langle j,k\rangle,s,s^\prime} \hat{\bm{e}}_z \cdot (\bm{\sigma}_{s,s^\prime} \times \bm{d}_{j,k}) c_{j,s}^\dagger c_{k,s^\prime}
\end{aligned}
\label{eq:r_hamiltonian}
\end{equation}
and,
\begin{equation}
\begin{aligned}
    H_\text{M} &= \lambda_{\text{ex}} \sum_{i,s,s^\prime} c_{i,s}^\dagger c_{j,s^\prime} [{\bm{n}} \cdot \bm{\sigma}]_{ss^\prime}
\end{aligned}
\label{eq:m_hamiltonian}
\end{equation}

\begin{figure}
    \centering
    \includegraphics[width=0.7\linewidth]{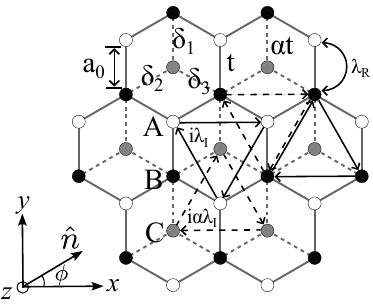}
    \caption{Schematic of $\alpha-\mathcal{T}_3$ lattice with sublattices A, B and C are denoted by white, black and grey circles, respectively. The model contains nearest-neighbour hopping $t$ (solid lines) and $\alpha t$ (dashe lines); next-nearest-neighbour intrinsic SOC $\lambda_I$ and $\alpha \lambda_I$ (solid and dashed arrows, respectively); spin-mixing Rashba spin-orbit coupling $\lambda_R$ (curved arrow). Magnetization orientation in real space is specified by $\hat{n}$.}
    \label{fig:schematic_lattice}
\end{figure}

Where $c_{i,s}^\dagger (c_{i,s})$ is the creation (annihilation) operator for an electron on lattice site $i$ and spin $s\;(\uparrow, \downarrow)$. Indices $i,j$ and $k$ belongs to sublattice A, B and C respectively and $\sigma=(\sigma_0, \sigma_x, \sigma_y, \sigma_z)$ comprises the Pauli matrices representing electron spin. The nearest neighbour hoppings are denoted in Fig.~\ref{fig:schematic_lattice}, are given by $\bm{\delta}_1=a_0 (0, 1)$, $\bm{\delta}_2 = a_0 (-\frac{\sqrt{3}}{2}, -\frac{1}{2})$ and $\bm{\delta}_3 = a_0 (-\frac{\sqrt{3}}{2}, \frac{1}{2})$, where $a_0$ is the length of nearest neighbour distance. The first two terms in Eq.~(\ref{eq:km_hamiltonian}) are nearest-neighbour (NN) hopping between the A and B sites with coupling strength $t$, and between B and C sites with coupling strength $\alpha t$. The third and fourth terms represent intrinsic spin-orbit coupling (SOC) that couples the same spins on next-nearest-neighbours (NNN), between A-A via B and B-B via A with hopping strength $\lambda_I$ and C-C via B and B-B via C with hopping strength $\alpha \lambda_I$ \cite{PhysRevB.103.075419}. 
$\nu_{i,j}$ ($\nu_{j,k}) = +1 (-1)$, denotes the anticlockwise (clockwise) direction of the NNN hopping path for spin-up (spin-down) from site $j(k)$ to $i(j)$. The NNN vectors are given by $\bm{\nu}_1 = (\bm{\delta}_2 - \bm{\delta}_3)$, $\bm{\nu}_2 = (\bm{\delta}_3 - \bm{\delta}_1)$ and $\bm{\nu}_3 = (\bm{\delta}_1 - \bm{\delta}_2)$. Eq.~\ref{eq:r_hamiltonian} describes Rashba SOC that mixes spins of nearest neighbours, with coupling strength $\lambda_R\;(\alpha\lambda_R)$, where $d_{i,j}\;(d_{j,k})$ represents a unit vector pointing from site $j(k)$ to site $i(j)$. The last term of Eq.\ref{eq:full_hamiltonian} displayed in Eq.\ref{eq:m_hamiltonian}, is the ferromagnetic exchange coupling of strength $\lambda_\text{ex}$ with N\'{e}el vector $\hat{n} = (\sin\theta \cos\phi, \sin\theta\sin\phi, \cos\theta )$, where $\phi$ is azimuthal angle and $\theta$ is polar angle in a spherical co-ordinate system, measured with respect to the positive $x$ and $z$ axis respectively.
The momentum space Hamiltonian in the sublattice basis $\{\ket{A_\uparrow},\ket{B_\uparrow},\ket{C_\uparrow},\ket{A_\downarrow},\ket{B_\downarrow},\ket{C_\downarrow} \}^T$ obtained by Fourier transforming Eq.\ref{eq:full_hamiltonian} can be described as,

\begin{equation}
    H(\boldsymbol{k}) = \begin{pmatrix}
        H_\uparrow & H_{\uparrow\downarrow}\\
        H_{\downarrow\uparrow} & H_\downarrow
    \end{pmatrix}
    \label{eq:H_k_full}
\end{equation}

Where, $H_\uparrow (H_\downarrow)$ is the spin-up (spin-down) Hamiltonian and $H_{\uparrow\downarrow} (H_{\downarrow\uparrow})$ is the spin-mixing part of the Hamiltonian.

\begin{equation}
\scalebox{0.7}{$
    H_{\uparrow(\downarrow)} = \begin{pmatrix}
        +(-)(\lambda_\text{ex}^\perp + f_0(\bm{k})\lambda_I^A) & f(\bm{k},t) & 0\\[6pt]
        f^*(\bm{k},t) & +(-)(\lambda_\text{ex}^\perp - (1-\alpha)\lambda_I^B) & \alpha f(\bm{k},t)\\[6pt]
        0 & \alpha f^*(\bm{k},t) & +(-)(\lambda_\text{ex}^\perp - \alpha f_0(\bm{k})\lambda_I^C)
    \end{pmatrix}
$}
\end{equation}

\begin{equation}
    H_{\uparrow\downarrow} = \begin{pmatrix}
        \lambda_\text{ex}^\parallel e^{-i\phi} & \rho(\bm{k}) & 0\\[4pt]
        -\rho(-\bm{k}) & \lambda_\text{ex}^\parallel e^{-i\phi} & -\alpha\rho(\bm{k})\\[2pt]
        0 & \alpha\rho(\bm{-k}) & \lambda_\text{ex}^\parallel e^{-i\phi}
    \end{pmatrix}
    \label{eq:spin_mixing_hamiltonian}
\end{equation}
and $H_{\downarrow\uparrow} = H_{\uparrow\downarrow}^\dagger$ where, $f(\bm{k},t) = f_x(\bm{k},t) - i f_y(\bm{k},t)$,
\begin{multline}
f_x(\bm{k},t) = t \bigg\{\cos \left(a_{0} k_y \right) \\
+ 2 \cos \left( \frac{\sqrt{3} a_{0} k_x}{2} \right)
\cos \left( \frac{a_{0} k_y}{2} \right)
\bigg\},
\end{multline}

\begin{multline}
    f_y(\bm{k},t) = t\bigg\{\sin \left(a_{0}k_y\right)\\
    - 2\cos \left( \frac{\sqrt{3} a_{0}k_x}{2} \right)
     \sin \left( \frac{a_{0}k_y}{2} \right)
     \bigg\},
\end{multline}

\begin{multline}
f_0(\bm{k},t) = 2 \bigg\{ \sin \left( \sqrt{3}a_{0} k_y \right)\\
- 2\sin \left( \frac{\sqrt{3} a_{0} k_x}{2} \right) \cos \left( \frac{3a_{0} k_y}{2} \right)\bigg\},
\end{multline}

\begin{multline}
    \rho(\bm{k}) = i\lambda_R \bigg\{ e^{-ia_0k_y}\; + \\
     \; 2 \; e^{\frac{ia_0k_y}{2}}\cos{\left( \frac{\sqrt{3}a_0k_x}{2} + \frac{2\pi}{3} \right)} \bigg\}
\end{multline}

\begin{multline}
    \rho(-\bm{k}) = i\lambda_R \bigg\{ e^{ia_0k_y} \; + \\
     2 \; e^{-\frac{ia_0k_y}{2}}\cos{\left( \frac{\sqrt{3}a_0k_x}{2} - \frac{2\pi}{3} \right)} \bigg\}
\end{multline}
We restrict the N\'{e}el vector to be either along the $z$-direction (denoted by $\lambda_\text{ex}^\perp$) or in the $xy$-plane (denoted by $\lambda_\text{ex}^\parallel$).

\section{Result and discussions}\label{sec:results}
We initially fixed the ferromagnetic exchange coupling term along the out-of-plane ($\hat{z}$) direction and looked into the bulk band gap without and with Rashba coupling. A direct band gap closing and reopening suggests a topological phase transition yields a change in topological invariants, whereas an indirect band gap closing leads to a metallic phase. We then looked into the edge band picture of the topological phases in a nanoribbon geometry. We later studied the in-plane magnetization case and showed how 1st-order topological edge bands gap out and a higher-order topology emerges in the same system. We kept $t=-1$, $a=1$ and $\hbar=1$ for convenience, and all other Hamiltonian parameters are in terms of $t$.
\subsection{Phase diagram}\label{sec:section_A}

    \begin{figure}[t]
    \centering
    \includegraphics[scale=0.35]{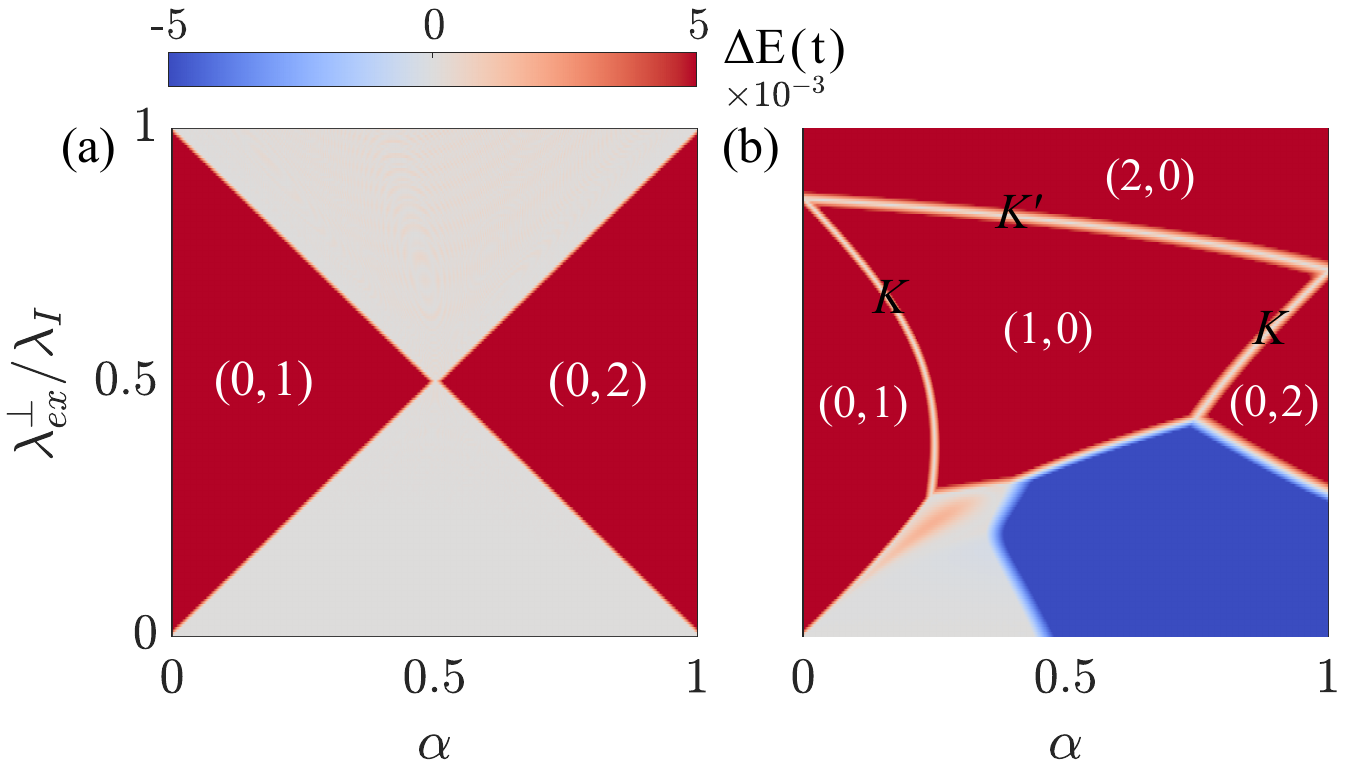}
    \caption{Phase diagram of the global bulk gap as a function of $\alpha$ and exchange coupling $\lambda_{ex}^\perp$ for out-of-plane magnetization for (a) $\lambda_R = 0$ and (a) $\lambda_R = 0.05$. The Chern numbers $(C, C_\sigma)$ are used to index the TI phases. A negative bandgap indicates transition to a metallic phase due to the indirect band gap closing as the band gaps at different k values overlap.}
    \label{fig:alpha_mz_bgap_phase}
\end{figure}

\begin{figure*}
    \centering
    \includegraphics[scale=0.33]{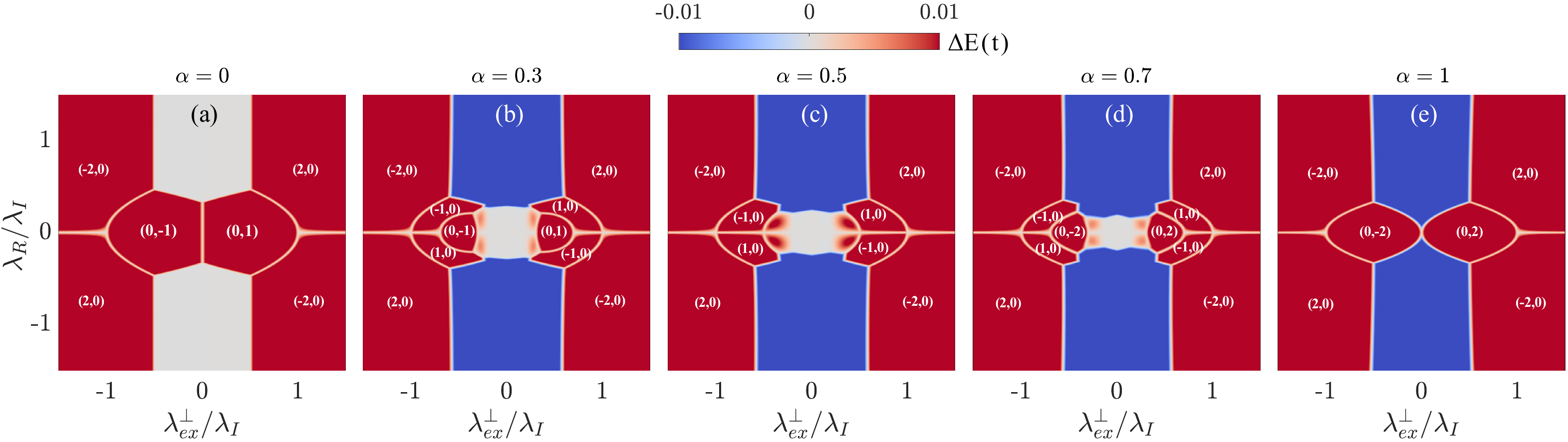}
    \caption{Phase diagram of bulk band gap in the $\lambda_R/ \lambda_I$ - $\lambda_{ex}^\perp/\lambda_I$ plane, for different values of $\alpha$. $(C, C_\sigma)$ indicates the total and projected spin Chern number of the occupied subspace for different topological phases. The zero band-gap lines indicate a phase transition, whereas the blue region suggests an indirect band-gap closure.}
    \label{fig:lambda_r_lambda_ex_phase}
\end{figure*}

Fig.\ref{fig:alpha_mz_bgap_phase}(a) and Fig.\ref{fig:alpha_mz_bgap_phase}(b) show the numerically calculated bulk band gap phase diagram as a function of out-of-plane exchange field $\lambda_{ex}^\perp$ and the lattice parameter $\alpha$,  in the absence and presence of Rashba SOC, respectively. The grey region denotes a direct band-gap closing, whereas the blue region indicates closing of the indirect bandgap, both representing a metallic phase. The insulating phases are indexed by the pair $(C, C_s)$, where $C$ is the total Chern number and $C_s$ is the projected spin Chern number of the occupied subspace. While the total Chern number ($C$) identifies the QAH topology, the QSH phase is characterised by the projected spin Chern number  $(C_s)$, as the exchange field explicitly breaks the TRS and hence the $Z_2$ invariant fails to characterise the QSH phase.\\

In systems where $\hat{s}_z$ commutes with the Hamiltonian, the Hilbert space can be decoupled into two independent sectors (spin-up and spin-down), and the spin Chern number is given by $C_s = (C_\uparrow - C_\downarrow)/2$, where the two individual Chern numbers can be obtained by integrating the Berry curvature of the relative spin-up and spin-down subspace. For the time-reversal symmetric case, the $Z_2$ invariant of the system is simply given by $C_s$ mod $2$. In the present model, however, Rashba SOC mixes the spin degree of freedom such that $[\hat{H}, \hat{S_z}] \neq 0$, preventing the decomposition of the Hilbert space into independent spin sectors. Consequently, the conventional spin Chern number is no longer an integer. \\
A spin Chern number for $s_z$ non-conserving system was initially introduced to characterise the topological order and to show the robustness of the QSH conductivity, but in a finite-sized system with certain boundary conditions \cite{PhysRevLett.97.036808, PhysRevLett.95.136602}.
In Ref \cite{PhysRevB.80.125327}, Prodan mathematically formalised and redefined the spin Chern number within the framework of band theory, as an alternative topological invariant to characterise the QSHI phase, which does not explicitly rely on the presence of time-reversal symmetry. The robustness of this invariant is due to two spectral gaps, the insulating gap of the Hamiltonian ($\Delta E$) and the eigenvalue spectrum of the valence-projected spin matrix ($\Delta{s_z}$), which is referred to as the spin gap. This quantity we refer to as the projected spin Chern number (PSCN). Remarkably, this strategy holds even if $[H, S_z]\neq 0$ as long as the operator displays a spin gap.  

In order to numerically compute the PSCN, we first construct the valance-band projector onto the occupied subspace, $P(k) = \sum\limits_{n \in occ} \ket{V_n(k)}\bra{V_n(k)}$, where $\ket{V_n(k)}$ denotes the Bloch eigenstate of the $n$\textsuperscript{th} occupied band~\cite{PhysRevB.80.125327, Lin2024}. The projected spin operator is then defined as $\tilde{S} = P(k) (\sigma_z\otimes I_3) P(k)$. Now by diagonalizing $\tilde{S}$, yields eigenvectors $\ket{\psi_v (k)}$ with corresponding eigenvalues $\epsilon_v(k)$. The occupied subspace can be separated into positive and negative-spin sectors corresponding to $\epsilon_v(k)>0$ and $\epsilon_v(k) <0$, respectively, provided that the spectrum of the projected spin operator remains gapped around zero throughout the Brillouin zone. One can define the projectors onto the positive and negative eigenspace of $\tilde{S}$ as, $P_+(k)=\sum\limits_{\epsilon_v(k)>0} \ket{\psi_v(k)}\bra{\psi_v(k)}$ and $P_-(k)=\sum\limits_{\epsilon_v(k)<0} \ket{\psi_v(k)}\bra{\psi_v(k)}$ respectively \cite{PhysRevB.111.035411}.
Using $P_\pm (k)$, one can assign an integer Chern number $C_\pm$ to positive and negative spin sectors, computed using the non-abelian Fukui method \cite{fukui2005chern} in momentum space. The PSCN is then defined by $C_s=(C_+ - C_-)/2$.

Instead of constructing the full projected spin operator in the original Hilbert space, one may equivalently define an effective projected spin operator within the occupied subspace, $S_{ij}(k) = \bra{V_i(k)}(\sigma_z \otimes I_3)\ket{V_j(k)}$, with $i,j = 1,2,....N_\mathrm{occ}$, which forms an $N_\mathrm{occ}\times N_\mathrm{occ}$ Hermitian matrix. The eigenvectors obtained from diagonalising $S(k)$ corresponding to positive and negative eigenvalues can similarly be used to calculate $C_+$ and $C_-$, and hence the PSCN, using the Fukui method \cite{saha2021eightfold, Lin2024}. 

In the absence of Rashba SOC, the phase diagram is much simpler, with two quantum spin Hall phases separated by metallic zero band-gap regions. Since the phase transitions occurs exclusively at the $\textbf{K}$-point, the phase boundaries can easily be obtained analytically to be $\alpha = \lambda_{ex}^\perp/\lambda_I$ and $\alpha = 1 - \lambda_{ex}^\perp/\lambda_I$ for $\lambda_{ex}^\perp/\lambda_I < 0.5$ and  $\lambda_{ex}^\perp/\lambda_I > 0.5$ respectively, whereas the other valley remains gapped throughout. These two phase boundaries merge at the critical point $\lambda_{ex}^\perp/\lambda_I = 0.5$. In the TRS preserving Kane-Mele $\alpha-\tau_3$ model, it is well established that the QSH phase undergoes a transition from $C_s = 1$ to $C_s =2$ at $\alpha=0.5$~\cite{PhysRevB.103.075419}. 
However, the phase transition happens between the spin-resolved middle band and the valence (conduction) band for the spin-up (spin-down) sector. In contrast, when an out-of-plane exchange field is introduced, the spin-resolved bands shift upward (spin-up) and downward (spin-down) in energy, and the transition between $C_s = 1$ to $C_s = 2$ QSH phases occurs at the Fermi level solely at the $\text{K}$-valley. A change in the direction of $\lambda_{ex}^\perp$ from $+\hat{z}$ to $-\hat{z}$ alters the valley from $\text{K}$ to $\text{K}^\prime$.

\hspace*{5pt} Upon introducing a finite Rashba SOC, the system is no longer fully spin-up or spin-down polarized, as the Rashba interaction mixes the in-plane spin component with momentum, thereby breaking $s_z$ conservation. Consequently, the phase boundaries of the QSH phases are modified, and two distinct QAH phases emerge, as shown in Fig.~\ref{fig:alpha_mz_bgap_phase}(b). Among them, the $(2,0)$ phase (that arises irrespective of $\alpha$-values) is the well-known QAH phase reported previously for both the Honeycomb and Dice lattices with Rashba SOC and ferromagnetic exchange coupling in the absence of intrinsic SOC\cite{PhysRevB.82.161414,4sq5-x2jy}. More remarkably, our model hosts an additional QAH phase with Chern number $C=1$, labelled as $(1,0)$, which appears exclusively for intermediate values of the lattice parameter ($0<\alpha<1$).
The phase transition from QSH $(0,1)$ to QAH $(1,0)$, and QAH $(1,0)$ to QSH $(0,2)$ are accompanied by a bulk band closing and reopening at $\text{K}$ valley, whereas transition from QAH $(1,0)$ to QAH $(2,0)$ happens at $\text{K}^\prime$ valley.
These phase boundaries can be obtained analytically from the energy spectrum at the valley points. The energy eigenvalues at $\text{K}(\text{K}^\prime)$ are given by $E_\eta^1 = -(\lambda_I+\eta \lambda_\text{ex}^\perp)$, $E_\eta^2 = -(\alpha\lambda_I-\eta\lambda_{ex}^\perp)$, $E_\eta^{3\pm} = \frac{1}{2}\Big\{ (2-\alpha)\lambda_I \pm\sqrt{(\alpha\eta\lambda_I+2\lambda_{ex}^\perp)^2 + 16\lambda_R^2} \Big\}$ and $E_\eta^{4\pm} = \frac{1}{2}\Big\{ (2\alpha-1)\lambda_I\pm\sqrt{(\lambda_I-2\eta\lambda_{ex}^\perp)^2+16\alpha^2\lambda_R^2} \Big\}$, where $\eta = +(-)$ for $\text{K}(\text{K}^\prime)$. 

The critical exchange strength is obtained from the condition that the highest occupied band touches the lowest unoccupied band at $\text{K}$ or $\text{K}^\prime$ point. Accordingly, the phase boundaries between QSH $(0,1)$ and QAH $(1,0)$, QAH $(1,0)$ and QSH $(0,2)$ and QAH $(1,0)$ to QAH $(2,0)$ are determined by the conditions $\Big|E_+^{3-} - E_+^{4+}\Big| = 0$, $\Big| E_+^2 - E_+^{4-} \Big| = 0$, and $\Big|E_-^{3-} - E_-^{1}\Big|=0$ respectively.
It is to note that, although the bulk energy gap remains finite throughout the QAH phase, the spin gap closes, rendering the $C_+$ and $C_-$, and hence PSCN $C_s$, ill-defined. Therefore, the QAH phases are characterised solely by the total (charge) Chern number, which is computed using the Fukui-Hatsugai-Suzuki method on the occupied subspace of the energy spectrum\cite{fukui2005chern}. The spin spectrum of all four distinct phases, QSH $(0,1)$, QSH $(0,2)$, QAH $(1,0)$ and QAH $(2,0)$ are shown in Appendix~\ref{app:spin_spectrum}.

\begin{figure}
    \centering
    \includegraphics[scale=0.45]{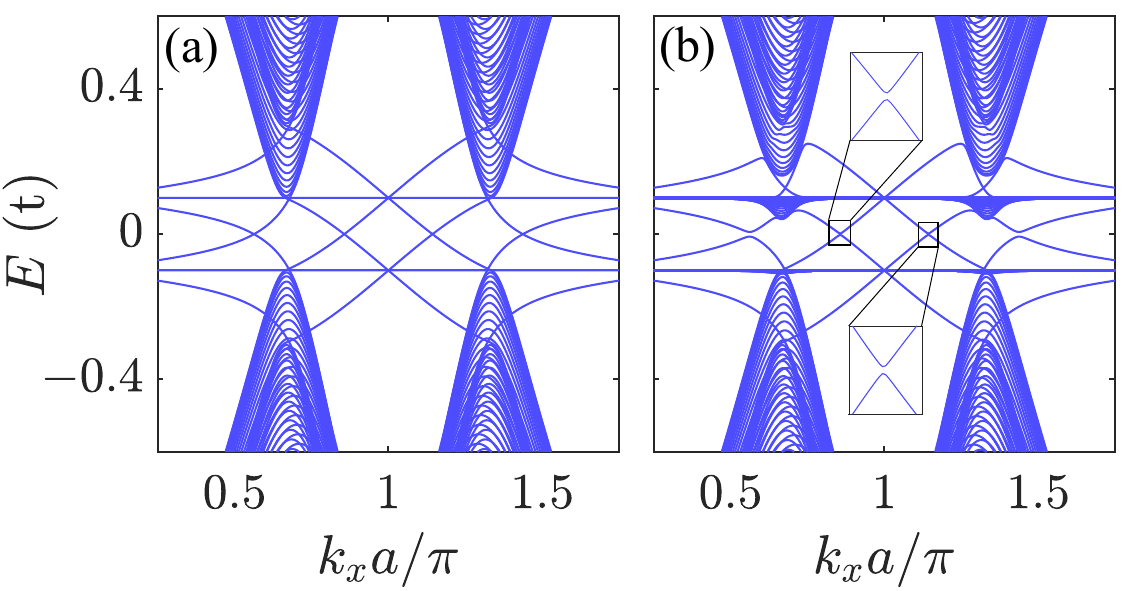}
    \caption{Electronic spectrum of zig-zag edged $\alpha-\tau_3$ ribbon for $\alpha=1$. The parameters are chosen to be $\lambda_I = 0.2$, $\lambda_{ex}^\perp = 0.1$ and $\lambda_R = 0$ (a) and $\lambda_R = 0.02$ (b). The inset shows gapped edge bands in the presence of Rashba coupling. The width of the ribbon considered is $30$nm, containing $141$ zigzag chains along the width.}
    \label{fig:ebs_rashba_comparison}
\end{figure}

\begin{figure}[!htb]
    \centering
    \includegraphics[scale=0.43]{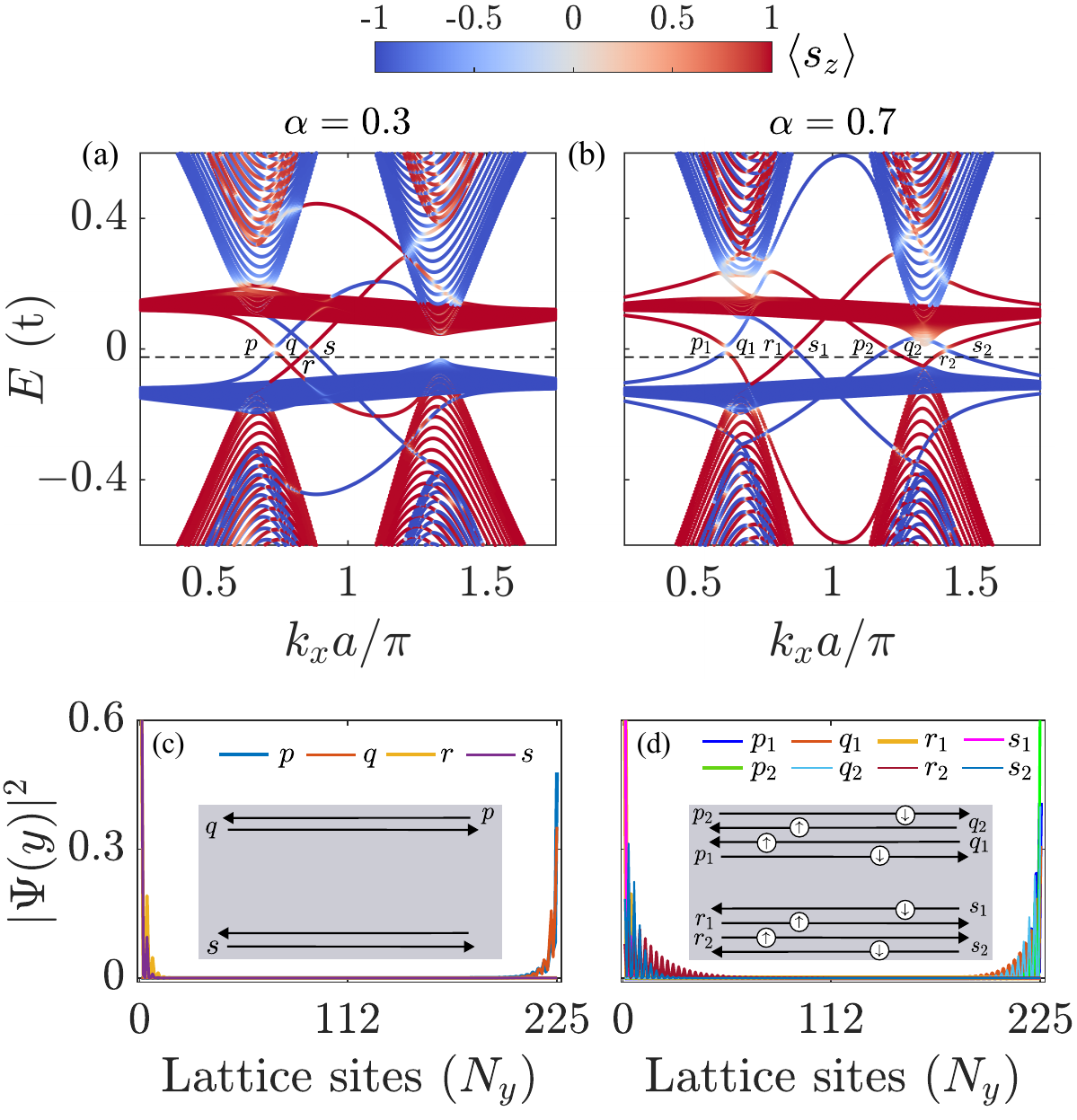}
    \caption{(upper panel) Calculated band structure of TRS broken zig-zag $\alpha-\tau_3$ nanoribbon in QSH phase for (a1) $\alpha=0.3$ and (a2) $\alpha=0.7$. The colour scale represents the spin expectation value $\langle s_z\rangle$. At a given Fermi energy ($E_f = -0.025$) in the bulk gap, there exist four edge modes for $\alpha =  0.3$, labeled as $p$, $q$, $r$ and $s$, and eight edge modes for $\alpha=0.7$ labeled $p_1$, $q_1$, $r_1$, $s_1$, $p_2$, $q_2$, $r_2$ and $s_2$. (lower panel) Probability density distribution with the lattice sites along the finite direction (b1) and (b2), corresponding to edge modes shown in (a1) and (a2), respectively. The insets schematically illustrate the edge modes localization and helicity on a rectangular slab, the arrows indicating the spin-polarization. The zigzag nanoribbon has a width of $16$ nm, containing $N_y = 225$ ABC sites. We use $\lambda_I = 0.25$, $\lambda_{ex}^\perp$ = 0.12 and $\lambda_R = 0.02$.}
    \label{fig:ebs_pd_qsh}
\end{figure}

To explore the evolution of the topological phases with Rashba SOC, Fig.~\ref{fig:lambda_r_lambda_ex_phase}(a) - Fig.~\ref{fig:lambda_r_lambda_ex_phase}(e) show the bulk-gap phase diagram in the parameter space of $\lambda_\text{ex}^\perp/\lambda_I$ and $\lambda_R/\lambda_I$ for different values of $\alpha$. Since the spinful $\alpha-\mathcal{T}_3$ model consists of six bands, the bulk gap is defined as the difference between band-3 (the highest occupied band) and band-4 (the lowest unoccupied band). It should be noted that the phase boundaries for $\alpha=0$ do not quantitatively coincide with those of the Honeycomb lattice reported in Ref.\cite{PhysRevLett.107.066602} for low exchange field strength. The discrepancy originates from the fact that, at $\alpha=0$ the $C$ sublattice is completely isolated, giving rise to two completely non-dispersive bands with energy splitting proportional to $\lambda_\text{ex}^\perp$. As long as the bulk gap is determined by these flat bands, the phase boundaries deviate from those of the honeycomb lattice. However, the phase diagram recovers the description when both associated energy bands are dispersive, following interband crossings between band-2 (band-4) and band-3 (band-5) in the valence (conduction) band. Nevertheless, the topological characterization remains identical because the isolated flat bands carry zero Berry curvature and hence do not contribute to the topological invariants. 
Several general features are evident from these phase diagrams. In the absence of Rashba SOC ($\lambda_R=0$), only spin Hall phases persist, irrespective of $\alpha$, and the phase boundary collapses to a single critical point at $\alpha=0.5$, consistent with Fig.\ref{fig:alpha_mz_bgap_phase}(a). Furthermore, the QAH phase with $C=\pm1$ is absent at the two limiting cases, $\alpha=0$ and $\alpha=1$, in agreement with the $\alpha$-$\lambda_{ex}^{\perp}$ phase diagram shown in (Fig.\ref{fig:alpha_mz_bgap_phase}(b)). This further confirms that the $C=\pm1$ QAH phase is an intrinsic feature of the intermediate $\alpha$ regime.

\subsection{Edge spectral properties} \label{sec:section_B}
To investigate the edge states in the TRS-broken QSH phase in the presence and absence of Rashba SOC, we calculate the energy spectrum of a quasi-one-dimensional zigzag edge of $\alpha-\tau_3$ nanoribbon, by considering periodic boundary conditions along the $x$-direction (see Fig.\ref{fig:schematic_lattice}) and finite termination along the $y$-direction. The calculated spectra of the TRS-broken QSH phase $(0,2)$ for $\alpha=1$ are shown in Fig.\ref{fig:ebs_rashba_comparison}(a) for $\lambda_R=0$ and in Fig.\ref{fig:ebs_rashba_comparison}(b) for $\lambda_R=0.02$. The ribbon has a width of $30$nm, consisting $141$ zigzag chains ($423$ ABC sites). In contrast to the TRS-preserved QSH phase, where the helical edge states reside in the gaps between the middle and valence (conduction) bands for spin-up (spin-down) electrons and around the Fermi energy the bands are bulk bands \cite{PhysRevB.103.075419}, the TRS-broken QSH phase exhibits edge modes within the bulk gap around the Fermi energy. In the absence of Rashba SOC, the Hamiltonian retains a residual $U(1)$ spin-rotation symmetry about the $z$-axis, ensuring conservation of $S_z$. Consequently, counter-propagating edge modes with opposite spin-polarization cannot hybridize, despite the breaking of TRS by the exchange field, and the edge spectrum remains gapless. The inclusion of Rashba coupling breaks this residual symmetry, allowing spin mixing and inter-channel scattering between the counter-propagating edge modes, which opens a gap in the edge dispersion.

To examine how the helical edge spectrum evolves with $\alpha$, we calculate the spin-projected energy spectrum of a zigzag nanoribbon for two different $\alpha$-values, $\alpha=0.3\;(<0.5)$ and $\alpha=0.7\;(>0.5)$, as shown in Fig.~\ref{fig:ebs_pd_qsh}(a) and (b), respectively. The parameters are chosen such that the system remains in the QSH phase of the $\lambda_R/\lambda_I$ - $\lambda_\text{ex}^\perp/\lambda_I$ phase space (see Fig.~\ref{fig:lambda_r_lambda_ex_phase}(b) and Fig.~\ref{fig:lambda_r_lambda_ex_phase}(d)) for corresponding values of $\alpha$. One can easily distinguish the in-gap edge states from the bulk states. The small energy gap in the edge bands causes weakly dissipative spin transport, arising due to non-zero Rashba coupling as discussed earlier. The edge bands remain a dominant spin character over most of the Brillouin zone, showing nearly complete spin-polarization, with appreciable spin-mixing occurring only near the avoided crossings. Fig.~\ref{fig:ebs_pd_qsh}(a) exhibits a single pair of helical edge modes, whereas Fig.~\ref{fig:ebs_pd_qsh}(b) exhibits two pairs, in consistent with the PSCN calculated and labelled in $\lambda_R/\lambda_I$ - $\lambda_\text{ex}^\perp/\lambda_I$ phase diagrams.  The probability density distributions of the edge modes at the Fermi energy indicated by the black dashed line are plotted as a function of lattice sites along the finite direction in Fig.~\ref{fig:ebs_pd_qsh}(c) and Fig.~\ref{fig:ebs_pd_qsh}(d). The insets schematically illustrate the localization and helicity of the edge modes in a finite rectangular ribbon. We consider a nanoribbon of width $16$ nm, containing $75$ zigzag chains ($225$ ABC sites) for this calculation.    

\begin{figure}
    \centering
    \includegraphics[scale=0.5]{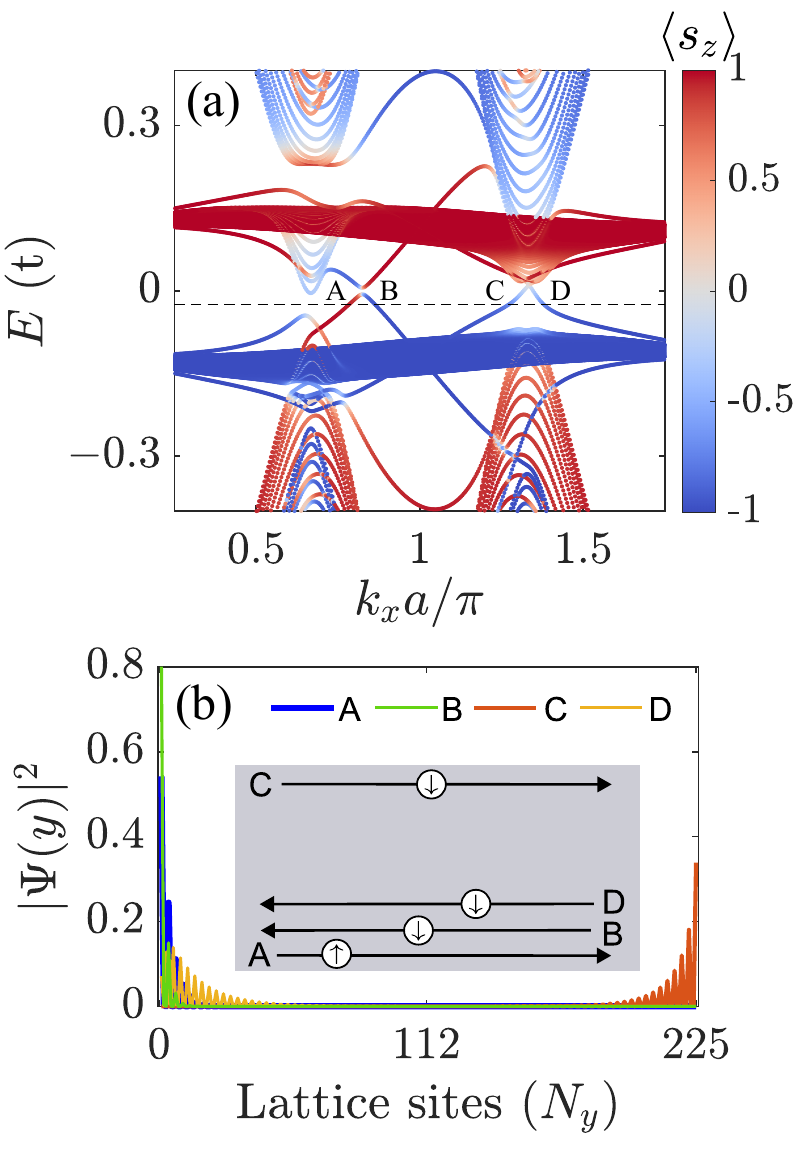}
    \caption{(a) Energy spectrum of zigzag $\alpha-\tau_3$ nanoribbon in the QAH(1,0) region of the phase space for $\alpha=0.5$. The colour represents the $s_z$-expectation value. A, B, C and D indicate the edge modes at a given Fermi energy (black dashed line). (b) Probability density distribution with the lattice sites along the finite direction, corresponding to the edge modes. The inset shows schematics of the localization and direction of the edge modes along a rectangular slab. We use $\lambda_I=0.2$, $\lambda_\text{ex}^\perp=0.12$ and $\lambda_R =0.05$.}
    \label{fig:qah1_ebs_pd}
\end{figure}

In the quantum anomalous Hall phase with Chern number $C=1$ [see Fig.~\ref{fig:alpha_mz_bgap_phase}(b)], the edge spectrum exhibits multiple edge modes as shown in Fig.~\ref{fig:qah1_ebs_pd}, even though only a single chiral mode is required by the bulk topological invariant. In particular, we observe two apparent pairs of edge modes (A, B) and (C, D), with three modes (A, B, D) localized on one edge and a single mode (C) on the opposite edge. Among these, A and C propagate along $+x$, while B and D propagate along $-x$. 
The spin-resolved spectrum shows that mode A is positive spin-polarized, whereas B, C, and D are predominantly negatively spin-polarized. Despite the presence of multiple modes, the net chirality at a given edge remains unity, consistent with the bulk Chern number $C=1$. The counter-propagating modes A and B, which coexist on the same edge, are not protected by any symmetry and therefore hybridize, leading to a gap in the edge spectrum. In contrast, although C and D propagate in opposite directions, they are localized on opposite edges and do not hybridize. Consequently, this pair of modes remains gapless and accounts for the single robust chiral edge channel required by the bulk topology. 
The valley origin of these edge modes can be identified from their connection to the bulk bands: the pair (A, B) is associated with the $K^\prime$ valley, whereas the gapless pair (C, D) originates from the $ K$ valley. To substantiate this picture, we calculate the valley-resolved Chern numbers of the bulk bands and obtain $C_K=1$ and $C_{K^\prime}=0$, yielding the total Chern number $C= C_K + C_{K^\prime}=1$. This establishes that the nontrivial topology is contributed entirely by the $K$-valley and the $K^\prime$-valley remains topologically trivial.
It is worth noting that this valley-polarized QAH phase only exists for intermediate $\alpha$-values ($\alpha \neq 0,1$).

\begin{figure*}
    \centering
    \includegraphics[scale=0.4]{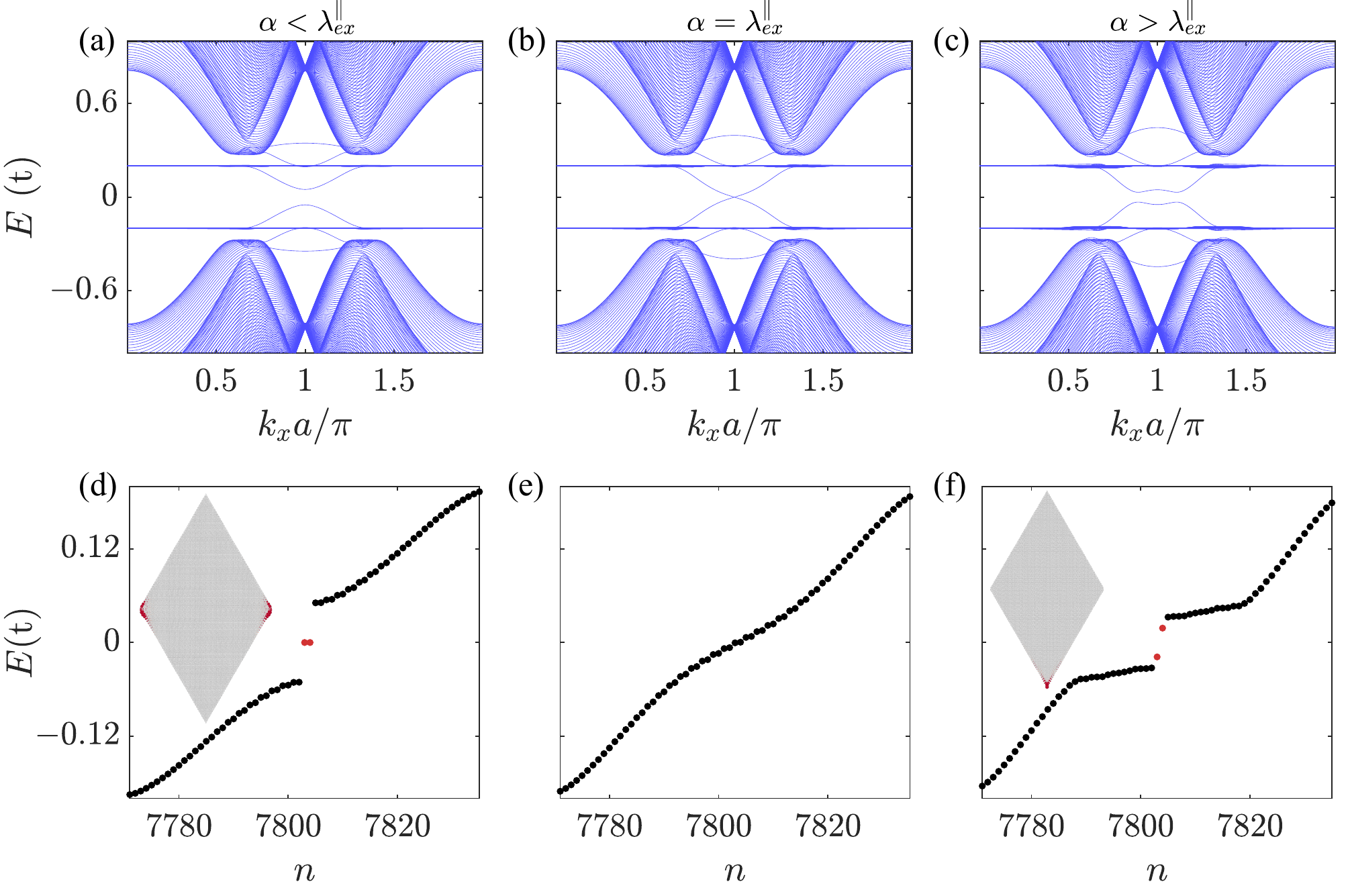}
    \caption{band structure of zigzag $\alpha-\tau_3$ nanoribbon with in-plane exchange field for (a) $\alpha = 0.15\:(< \lambda_{ex}^\parallel)$, (b) $\alpha = 0.2\:(=\lambda_{ex}^\parallel)$, and (c) $\alpha = 0.25\:(> \lambda_{ex}^\parallel)$. (d)-(f) Eigen-energies near zero energy of rhombic-supercell geometry corresponding to the same values of $\alpha$, highlighting in-gap corner modes in red circles. The insets display the spatial probability distributions of the states labelled by red circles. The other parameters are chosen to be $\lambda_I = 0.3$, $\lambda_{ex}^\parallel =  0.2$, $\lambda_R = 0$ and $\phi = 0$.}
    \label{fig:mx_soti_ebs}
\end{figure*}

\subsection{In-plane magnetization : Emergent second-order topological phase}\label{sec:section_C}
To investigate the effect of TRS breaking in the Kane - Mele $\alpha-\tau_3$ lattice, we also explored exchange fields with in-plane magnetization orientation at an arbitrary azimuthal angle and observed how the emergent phases evolve with the lattice parameter $\alpha$, in-plane exchange strength $\lambda_\text{ex}^\parallel$ and the strength of Rashba SOC $\lambda_R$. In-plane magnetization is well known for gapping out the first-order topological edge modes and generating second-order topological corner modes in the 2D lattice. Such behaviour was first demonstrated in the Kane-Mele model by Ren et al.\cite{PhysRevLett.124.166804} and has subsequently been reported in several other two-dimensional systems, such as strained Honeycomb \cite{PhysRevB.109.115424, Lahiri2024}, periodically driven isotropic and anisotropic ernevig-Hughes-Zhang (BHZ) model \cite{PhysRevB.100.115403}, Antiferromagnetic heterostructure \cite{PhysRevB.108.075401}, s-wave superconductor\cite{PhysRevB.107.085407}, graphene nanoflake with in-plane edge magnetization\cite{PhysRevB.106.165422}. 
The general physical mechanism is attributed to the formation of the Dirac mass domain walls, where adjacent edges acquire effective masses of opposite sign, giving rise to localized corner modes at their intersections \cite{PhysRevB.109.205417, PhysRevB.107.085407}. 

We find that a nonzero in-plane magnetization also gaps out the helical edge modes of the Kane-Mele $\alpha-\mathcal{T}_3$ Hamiltonian, as shown in Fig.~\ref{fig:mx_soti_ebs}(a). This in-plane magnetization yields edge-dependent behaviour, while gapping out the edge bands of the first-order topological insulator at the zigzag boundary, the conducting edge modes in the armchair-edge nanoribbon remain intact. 
To determine whether the resulting phase is a 2nd-order TI, we construct a rhombic supercell with a zigzag edge. The in-plane magnetization induces a pair of degenerate higher-order (2nd-order) topological corner modes at zero energy, each localized at the two obtuse corners of the finite nanoflake as shown in Fig.\ref{fig:mx_soti_ebs}(d). We have taken the in-plane magnetization along the $x$-direction; rotating the magnetization direction within the plane for any arbitrary $\phi$ does not qualitatively alter the SOTI modes.

\begin{figure*}
    \centering
    \includegraphics[scale=0.4]{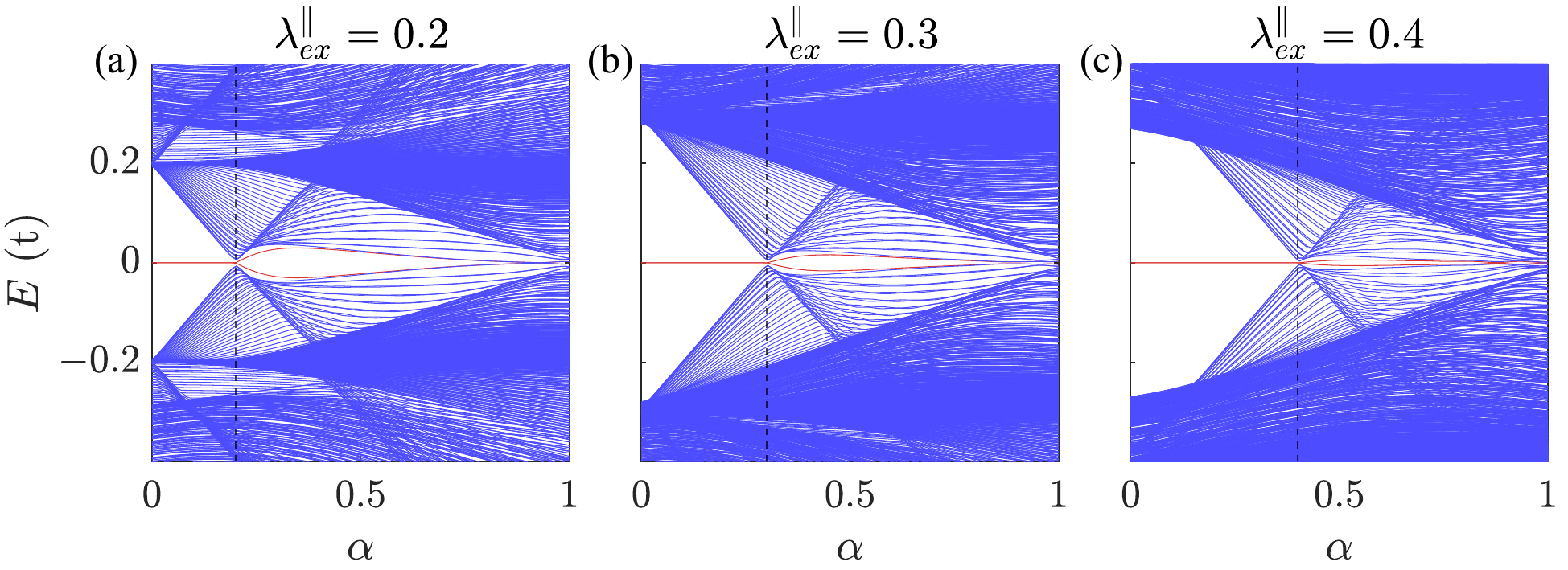}
    \caption{The real-space energy spectra of Kane-Mele-Rashba $\alpha-\mathcal{T}_3$ Hamiltonian in the presence of in-plane exchange field, projected on a rhombic supercell, plotted as a function of $\alpha$ for (a) $\lambda_{ex}^\parallel = 0.2$, (b) $\lambda_{ex}^\parallel = 0.3$, and (c) $\lambda_{ex}^\parallel = 0.4$. The black dashed lines indicate $\alpha = \lambda_{ex}^\parallel$, and the red lines suggest the SOTI modes. Here $\lambda_{I} = 0.3$ kept fixed.}
    \label{fig:mx_soti_spectra}
\end{figure*}

These zero-energy degenerate corner modes are highly tunable with the parameter $\alpha$. Specifically, these modes persist for $\alpha<\lambda_{ex}^\parallel$. We observe that for $\alpha>\lambda_{ex}^\parallel$, these zero-energy corner modes split into two higher-energy modes symmetrically away from the zero-energy. These higher-energy corner modes are localized at an acute corner of the finite nanoflake. The transition occurs via an edge band crossing of the zigzag boundary at $\alpha=\lambda_{ex}^\parallel$, with the bulk remaining fully gapped throughout.
This behaviour is illustrated in Figs.~\ref{fig:mx_soti_ebs}(a)-(c), which shows the edge band structures of a zigzag nanoribbon for $\alpha=0.15\;(< \lambda_{ex}^\parallel)$, $\alpha=0.2\;(= \lambda_{ex}^\parallel)$ and $\alpha=0.25\;(> \lambda_{ex}^\parallel)$ respectively. At $\alpha=\lambda_{ex}^\parallel$, the edge bands touch at a Dirac point (one-dimensional Dirac cone), then gap out again for $\alpha> \lambda_{ex}^\parallel$, signalling the phase transition.
Figs.~\ref{fig:mx_soti_ebs}(d)-(f) present the corresponding eigenenergies of a rhombic-supercell structure for the same $\alpha$ values. The insets display the spatial probability distribution of the states marked by red circles, clearly showing different corner localizations in the two SOTI phases. 
In Fig.~\ref{fig:mx_soti_spectra} (a) - (c), we plot the real-space eigenenergy spectrum of the rhombic supercell as a function of $\alpha$ for different strengths of the in-plane exchange field. The spectra clearly show that the zero-energy corner modes split into two higher-energy corner modes through the SOTI-to-SOTI phase transition exactly at $\alpha=\lambda_{ex}^\parallel$, and as $\alpha$ is further increased, the higher-energy modes approach and eventually merge with the bulk modes. Furthermore, an increase in field strength increases (decreases) the bulk gap at the zero-energy (higher-energy) in-gap corner modes.
We note that Rashba coupling is set to zero in these calculations, but the observed phenomena also persist even if a small Rashba coupling is induced, indicating that the SOTI phase arises solely due to the in-plane exchange field. 
Although the bulk gap remains finite throughout the SOTI phase transition and only closes at $\alpha=1$, the spin spectrum obtained by considering three occupied bands remains gapless irrespective of $\alpha$. However, for $1/3$ filling, when only two bands are occupied, the spin spectrum shows a clear gap throughout the Brillouin zone [see Fig.~\ref{fig:spin_spectrum_mx} in Appendix \ref{app:spin_spectrum}], and the calculated PSCN undergoes a transition from $C_\sigma =1$ to $C_\sigma=2$ at $\alpha=0.5$. This behaviour is consistent with the previous studies showing that an in-plane exchange field primarily gaps the helical edge states without destroying the underlying bulk topology inherited from the time-reversal symmetric Kane-Mele phase \cite{PhysRevB.110.235432, PhysRevLett.124.166804}. 

\section{conclusion}\label{sec:conclusion}
We considered a Kane-Mele-type quantum spin Hall pseudospin-1 $\alpha-\mathcal{T}_3$ system with time-reversal symmetry broken by an exchange field and $S_z$-conservation breaking Rashba coupling. 
We used the valence-projected spin operator method to calculate the projected spin-Chern number for identifying the QSH phases, since the $Z_2$ invariant is no longer well defined once time-reversal symmetry is broken.
Unlike systems with only two occupied bands, we found that for the present model with three occupied bands, this method works well in the QSH regime but fails in the QAH regime, as the spin-gap closes, rendering the positive and negative spin projectors ill-defined.  
For magnetization along the $z$-direction, we found two distinct QAH phases along with the two QSH phases (which are intrinsic properties of the TRS-preserved Kane-Mele $\alpha-\mathcal{T}_3$). We further identified the QAH $(1,0)$ phase as a valley-polarized QAH phase where the $K$-valley is topological and the $K^\prime$-valley remains trivial. This valley-imbalanced behaviour is only observed for $\alpha\neq0,1$. The helical edge modes in the QSH phases, however, are not completely dissipationless due to the hybridization between two counter-propagating opposite-spin channels localized on the same edges in the absence of $S_z$-conservation. In contrast, the chiral edge modes corresponding to the QAH phases are found to be gapless.  
For an exchange field along the in-plane direction, the system realizes a second-order topological insulator with zero-energy corner modes while preserving the underlying $\alpha$-dependent QSH topology inherited from the TRS-preserved Kane-Mele model. 
The system also shows an SOTI-to-SOTI phase transition at $\alpha=\lambda_{ex}^\parallel$, mediated solely by a gap closing of the zigzag edge bands while the bulk gap remains finite.
Our results motivate the investigation of spin-resolved topology in other two-dimensional systems with higher pseudospin. This work also motivates exploring projected internal degrees of freedom beyond spin, such as layer or orbital pseudospin, which may provide a general route for defining sector-resolved topological invariants in multiband systems.

\begin{acknowledgments}
We acknowledge the support provided by the Kepler Computing facility, maintained by the Department of Physical Sciences, IISER Kolkata, for various computational needs. P.P. acknowledges support from the Council of Scientific and Industrial Research (CSIR), India, for the doctoral fellowship. B.L.C. acknowledges the SERB with grant no. SRG/2022/001102 and “IISER Kolkata Start-up-Grant” Ref. No. IISERK/DoRD/SUG/BC/2021-22/376.
\end{acknowledgments}

\appendix
\renewcommand\thefigure{\thesection\arabic{figure}} 
\setcounter{figure}{0}
\section{Spin spectrum}\label{app:spin_spectrum}
\begin{figure}
    \centering
    \includegraphics[scale=0.4]{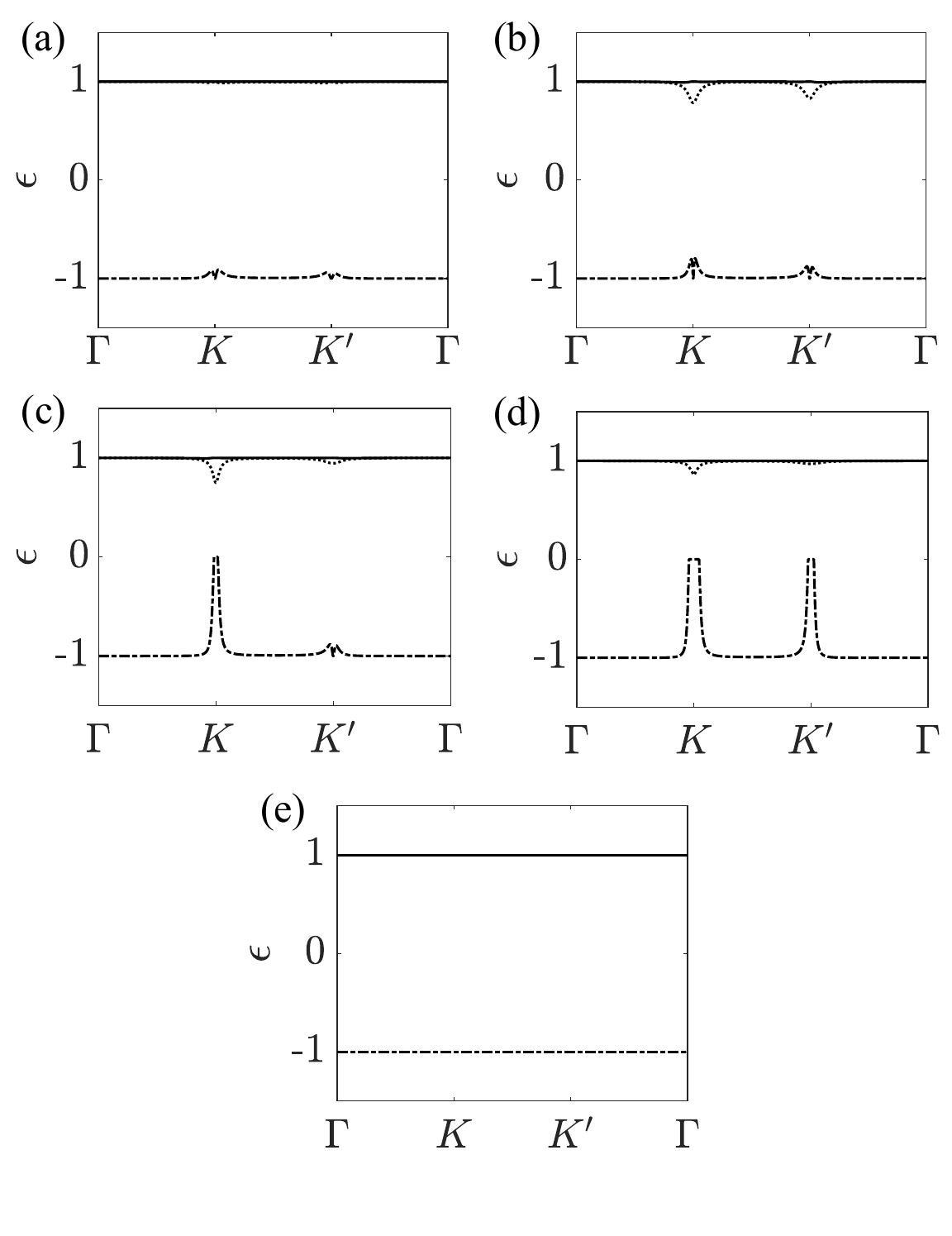}
    \caption{Eigenvalues of the projected spin operator $\tilde{S}$ along the high-symmetry path of the Brillouin zone for four distinct topological phases in Fig.\ref{fig:alpha_mz_bgap_phase}(b): (a) QSH $(0,1)$, (b) QSH $(0,2)$, QAH $(1,0)$ and QAH $(2,0)$. The corresponding parameters are (a) $\lambda_{ex}^\perp=0.1$, $\alpha=0.1$; (b) $\lambda_{ex}^\perp=0.1$, $\alpha=0.9$; (c)$\lambda_{ex}^\perp=0.12$, $\alpha=0.5$; and (d) $\lambda_{ex}^\perp=0.2$, $\alpha=0.5$. In (a)-(d), the remaining parameters are fixed at $\lambda_I=0.2$ and $\lambda_R=0.05$. (e) Eigenvalues of $\tilde{S}$ for the QSH $(0,2)$ phase in Fig.\ref{fig:alpha_mz_bgap_phase}(a), shown for comparison with the $\lambda_R=0$ case ($\lambda_I=0.2$, $\lambda_{ex}^\perp=0.1$, $\alpha=1$).}
    \label{fig:spin_spectrum_mz}
\end{figure}

\begin{figure}
    \centering
    \includegraphics[scale=0.4]{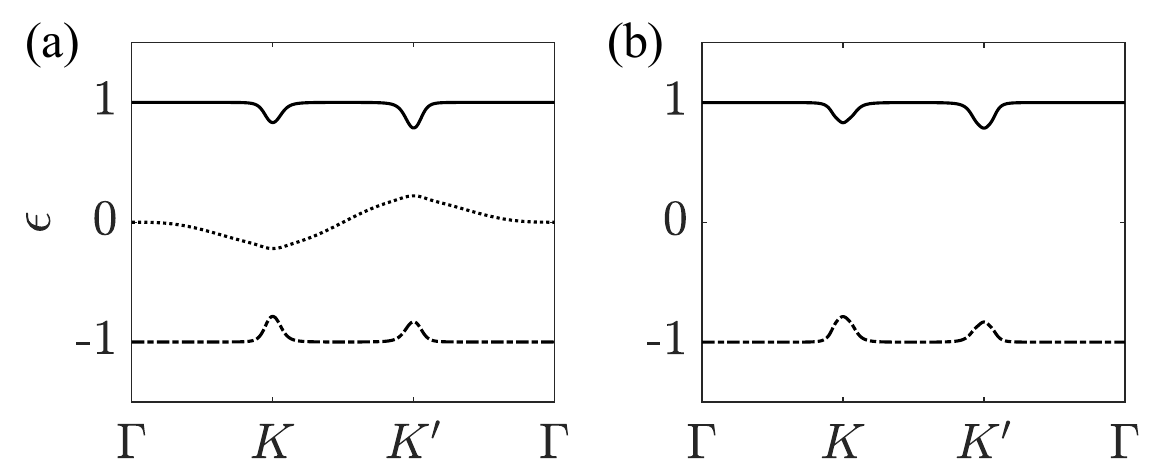}
    \caption{Eigenvalues of the effective projected spin operator $S$ along the high-symmetry path of the Brillouin zone in the SOTI phase, considering (a) three bands and (b) two bands in the occupied subspace. The parameters are fixed at $\lambda_I=0.3$, $\lambda_{ex}^\parallel=0.2$, $\lambda_R=0$ and $\alpha=0.15$.}
    \label{fig:spin_spectrum_mx}
\end{figure}

The diagonalization of $\tilde{S}$ described in Sec.~\ref{sec:section_A} yields six eigenvalues $\epsilon_v$ ($v \in [1,6]$), among them $N_\mathrm{occ}$ number of eigenvalues are non-zero ($\epsilon_1, \epsilon_2$, and $\epsilon_6$) and remaining are zero. We will ignore the zero eigenvalues which arise from states in the unoccupied subspace and provide no information. One can diagonalize the effective projected spin operator $S(k)$, discussed in \ref{sec:section_A}, to determine the spin spectrum, to avoid these additional zero eigenvalues.
The spin gap at each crystal momentum $k$ is defined as $\Delta S_z (k) \equiv \min\limits_v|\epsilon_v(k)|$, is the eigenvalue of $\tilde{S}$ with smallest absolute value \cite{Lin2024}. A non-zero min spin-gap, $\Delta S_z ^\text{min} = \min\limits_k \min\limits_v |\epsilon_v(k)|$ is required for the separation into $P_+$ and $P_-$ to be well defined \cite{yang2026layer}. In the presence of $s_z$-rotation symmetry, the non-zero eigenvalues are fixed at $\pm1$ and distributed symmetrically around zero as shown in Fig.~\ref{fig:spin_spectrum_mz}(e). If $s_z$-rotational symmetry-violating Rashba SOC is introduced, the eigenvalues of $\tilde{S}$ adiabatically deviate from $\pm1$ \cite{PhysRevB.110.214211}. In the QSH phases given in Fig.~\ref{fig:alpha_mz_bgap_phase}(b), the positive and negative spin eigenvalues are distributed around zero within the interval $[-1,1]$, as shown in Fig.~\ref{fig:spin_spectrum_mz}(a) and Fig.~\ref{fig:spin_spectrum_mz}(b). In the two QAH phases, spin eigenvalues of the negative eigenspace approach zero in the vicinity of the K (K$^\prime$) points [Figs.~\ref{fig:spin_spectrum_mz}(c) and Figs.~\ref{fig:spin_spectrum_mz}(d)], leading to the closing of the spin gap and rendering the PSCN ill-defined in this parameter region. 
Figure~\ref{fig:spin_spectrum_mx}(a) presents the spin spectrum corresponding to the SOTI phase shown in Fig.~\ref{fig:mx_soti_ebs}(d), with three occupied bands ($N_\mathrm{occ}=3$), i.e., the Fermi level lying in the bulk energy gap. Although the bulk gap remains finite ( $\Delta E \neq 0$ ), the spin gap vanishes ($\Delta S_0$), and PSCN can not be determined. However, for Fermi energy in the gap between band-2 and band-3 $(N_\mathrm{occ}=2)$, the corresponding spin-spectrum is gapped, as shown in Fig.~\ref{fig:spin_spectrum_mx}(b).

\bibliography{apssamp}

\end{document}